\documentclass[conference]{IEEEtran}
\IEEEoverridecommandlockouts

\usepackage{cite}
\usepackage{amsmath,amssymb,amsfonts}
\usepackage{algorithmic}
\usepackage{graphicx}
\usepackage{textcomp}
\usepackage{booktabs}
\usepackage{xcolor}
\usepackage{comment}
\usepackage{tikz}

\newcommand{\blackcircle}[1]{\tikz[baseline=(char.base)]{
            \node[shape=circle,fill=black,text=white,inner sep=1pt] (char) {#1};}}

\begin{document}

\title{Reimagining Memory Access for LLM Inference: Compression-Aware Memory Controller Design\\[-5pt]}

\author{
\centering
\IEEEauthorblockN{Rui Xie$^{1}$, Asad Ul Haq$^{1}$, Linsen Ma$^{1}$, Yunhua Fang$^{1}$, Zirak Burzin Engineer$^{2}$, Liu Liu$^{1}$, and Tong Zhang$^{1}$}
\IEEEauthorblockA{$^1$Rensselaer Polytechnic Institute, Troy, NY, USA \\
$^2$Wiseburn Da Vinci Science, El Segundo, CA, USA}
}

\maketitle

\begin{abstract}
The efficiency of Large Language Model~(LLM) inference is often constrained by substantial memory bandwidth and capacity demands. Existing techniques, such as pruning, quantization, and mixture of experts/depth, reduce memory capacity and/or bandwidth consumption at the cost of slight degradation in inference quality. This paper introduces a design solution that further alleviates memory bottlenecks by enhancing the on-chip memory controller in AI accelerators to achieve two main objectives: (1) significantly reducing memory capacity and bandwidth usage through lossless block compression~(e.g., LZ4 and ZSTD) of model weights and key-value (KV) cache without compromising inference quality, and (2) enabling memory bandwidth and energy consumption to scale proportionally with context-dependent dynamic quantization. These goals are accomplished by equipping the on-chip memory controller with mechanisms to improve fine-grained bit-level accessibility and compressibility of weights and KV cache through LLM-aware configuration of in-memory placement and representation. Experimental results on publicly available LLMs demonstrate the effectiveness of this approach, showing memory footprint reductions of 25.2\% for model weights and 46.9\% for KV cache. In addition, our hardware simulation results at 2\,GHz and 32 lanes achieve up to 2\,TB/s throughput with a modest area overhead (under 5.69\,mm\(^2\)), which underscores the viability of LLM-aware memory control as a key to efficient large-scale inference.

\end{abstract}

\section{Introduction}\label{sec:introduction}

Large Language Models (LLMs) have pushed the boundaries of natural language processing and generative AI by achieving impressive performance on tasks such as text generation, question answering, and code completion. As these models scale to billions or even trillions of parameters, however, their inference demands become increasingly constrained by memory bandwidth and capacity~\cite{alizadeh2023llm}. While the raw compute may be adequate, the sheer volume of parameters and associated data transfers often stalls the accelerator, resulting in suboptimal throughput and higher operational costs. 
Under these conditions, memory pressure intensifies, since the system must fetch billions of parameters from off-chip DRAM repeatedly, layer by layer, for each user request. Even small inefficiencies in memory access can cascade into significant overall slowdowns, magnifying both inference time and total operational cost. Furthermore, since LLMs are often fine-tuned or adapted for specialized domains, their parameters can be heterogeneous, featuring layers that might be more compressible than others. Such diversity further underscores the importance of a memory subsystem that can flexibly and efficiently handle different data patterns.


To alleviate this bottleneck, researchers have extensively explored \emph{lossy compression} techniques, such as pruning, quantization, and mixture of experts/depth, which reduce memory consumption at the cost of minor (though sometimes noticeable) degradation in inference quality~\cite{liu2023deja, frantar2022gptq, raposo2024mixture}. These methods have shown promise for large-scale LLMs, but they necessarily introduce approximations. By contrast, \emph{lossless compression} could preserve exact numerical fidelity while still shrinking the model footprint; however, floating-point data generally exhibit low compressibility under standard algorithms (e.g., LZ4~\cite{lz4-link}, ZSTD~\cite{zstd-link}), thus limiting their efficacy in practice.

This paper introduces an approach to notably increase the lossless compressibility of model weights and key-value~(KV) cache by optimizing their in-memory bit-level placement and representation. Motivated by prior works on bit-plane data placement~\cite{kim2016bit, cavigelli2019ebpcextendedbitplanecompression,xie2024smartquant}, we propose two specific techniques: (1) {\it Bit-plane disaggregation}: Instead of storing all bits of each weight or KV cache element~(e.g., FP16 or FP8) contiguously, we organize data by keeping same-position bits together, effectively creating a bit-level in-memory column-store~\cite{stonebraker2018c, nes2012monetdb}. (2)~{\it Cross-token KV cache clustering and de-correlation}: Observing that KV cache elements on the same channel of adjacent tokens tend to show stronger bit-level correlation, we cluster these elements together in memory and apply simple de-correlation mechanisms~(e.g., subtraction) to reduce bit-level entropy. By applying these techniques to model weights and KV cache, we can largely enhance their lossless compressibility (especially for KV cache), enabling effective use of conventional lossless compression algorithms like LZ4 and ZSTD.

Additionally, recent studies~\cite{tang2024quest, zhang2023h2o} reveal that the importance of different weights and KV cache elements varies significantly based on runtime context during inference, opening the door for dynamic quantization to further improve efficiency. With AI accelerators now supporting variable-precision arithmetic at the hardware level, these systems can leverage context-dependent quantization~(e.g., dynamically adjusting precision from FP16 to FP8) to enhance computational efficiency. However, it is not immediately clear how memory bandwidth/energy consumption can be proportionally reduced. Coincidentally, our {\it bit-plane disaggregation} technique can fill this missing link: it enables memory bandwidth and energy consumption to scale gracefully with context-adaptive dynamic quantization of weights and KV cache.

To ensure practical feasibility, we propose integrating these techniques, along with hardware-accelerated lossless compression and decompression, into the on-chip memory controller of AI accelerators. This approach minimizes changes to the computational framework and software stack, as the memory controller only needs to be aware of the data semantics of weights and KV cache. All bit-level data manipulations and compression tasks occur within the memory controller, remaining transparent to the computing fabric and software stack. Through extensive experiments on publicly available LLMs, we demonstrate that our approach achieves significant reductions in memory footprint without compromising inference accuracy. Experimental results reveal memory footprint reductions of 25.2\% for model weights and 46.9\% for KV cache when using ZSTD lossless compression. Moreover, simulations conducted using DRAMSim3~\cite{li2020dramsim3} demonstrate that, compared to straightforward in-memory placement, our proposed approach can reduce data load latency by up to 30.0\% and lower memory access energy consumption by 29.9\%.

\section{Background and Motivation}\label{sec:background}

\subsection{Memory Challenges in LLM Inference}
\begin{figure}[htbp]
	\centering
    \includegraphics[width=\linewidth]{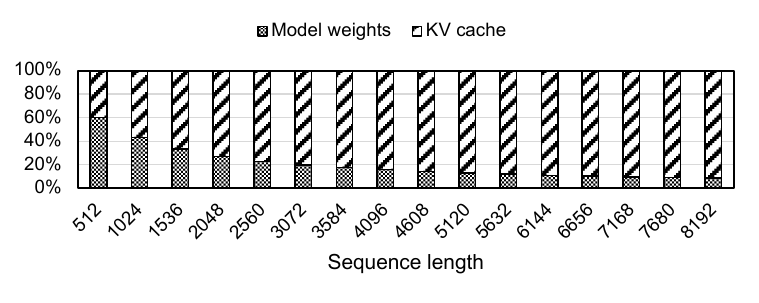}
	\caption{Percentage contribution of KV cache and model weights to total memory footprint with increasing sequence length in LLaMA 3.1 8B model.}
\label{fig.kv cache vs model}
\end{figure}

During LLM inference, the model repeatedly generates tokens by passing the entire input sequence (including previously generated tokens) through multiple layers. In large models (e.g., LLaMA 3.1 405B~\cite{meta2024llama}, DeepSeek R1 671B~\cite{deepseek2025r1}), storing these weights already demands significant capacity (750GB of LLaMA 3.1 405B, 1543GB of DeepSeek R1 671B). Beyond the parameters, every token update must also preserve a key-value (KV) cache that expands with sequence length. As shown in Fig.~\ref{fig.kv cache vs model}, once the sequence extends beyond a few thousand tokens, the KV cache can overshadow all other memory components, exceeding 90\% of the total footprint in the LLaMA 3.1 8B model. This quickly becomes a \textit{capacity crisis}, requiring substantial DRAM or HBM provisioning to prevent frequent swapping or out-of-memory failures.

In autoregressive decoding, each new token must read and compute against all model weights layer by layer, incurring large, repeated bandwidth demands. Moreover, tokens in the KV cache must be fetched (and updated) for attention calculations. Hence, both model weights and historical context collectively exert high bandwidth pressure on the memory subsystem. Even if the hardware provides ample compute throughput, inference throughput can stall unless memory traffic (i.e., read requests for weights and KV data) can be served rapidly. Alleviating this bandwidth bottleneck is often as critical as reducing raw model size, since partial or frequent memory stalls directly prolong token generation latency.

\subsection{Lossless compression on LLM}

Lossless compression holds the promise of reducing memory usage without degrading inference quality, making it an attractive complement to existing \emph{lossy} techniques. In principle, general-purpose methods such as LZ4~\cite{lz4-link} and ZSTD~\cite{zstd-link} can detect repeated byte sequences, zeros, or short runs of similar data, thus condensing the stored representation. However, as shown in Table~\ref{tab:direct compression ratios}, straightforwardly applying these compressors to LLM data (i.e., model weights and key-value (KV) caches) yields only limited success. LZ4 fails to reduce the footprint of weights and KV caches in most cases, and ZSTD attains moderate gains for some model parameters (e.g., 23\% on Gemma 2 2B weights~\cite{gemma2-link}).

\begin{table}[htbp]
\centering
\caption{Model weights and KV cache footprint reduction under lossless compression.}
\label{tab:direct compression ratios}
\resizebox{\columnwidth}{!}{%
\begin{tabular}{@{}lccccc@{}}
\toprule
\textbf{Comp. Method} & \textbf{LLaMA 3.1 8B} & \textbf{Gemma 2 2B} & \textbf{Mistral 7B} & \textbf{OPT 13B} & \textbf{Mixtral 8×7B} \\ \midrule
\multicolumn{6}{c}{\textbf{Model Weights}} \\ \midrule
LZ4 & 0.0\% & 11.5\% & 0.0\% & 0.0\% & 18.0\% \\
ZSTD & 20.6\% & 23.0\% & 17.3\% & 19.4\% & 21.3\% \\ \midrule
\multicolumn{6}{c}{\textbf{KV Cache on BookSum Dataset}} \\ \midrule
LZ4 & 0.0\% & 0.0\% & 0.0\% & 0.0\% & 0.0\% \\
ZSTD & 6.5\% & 2.9\% & 0.9\% & 2.0\% & 3.8\% \\ \bottomrule
\end{tabular}%
}
\end{table}

A major impediment is the floating-point data format, where exponent and mantissa bits are intermixed, leading to high byte-level entropy under conventional compression algorithms. The KV cache, which captures a dynamically evolving representation of token embeddings, is particularly challenging: its content shifts every time new tokens are processed, which offers fewer obvious redundancies. Nevertheless, recent studies~\cite{liu2024kivi,zhang2024kv,hooper2024kvquant} have highlighted potential opportunities for compression by observing that certain channels (or embedding dimensions) exhibit similar numeric patterns across consecutive tokens. For instance, in LLaMA 2, the relative reconstruction error—when grouping KV cache data by channel—can be significantly lower (e.g., 4.55) compared to grouping by token (13.67)~\cite{liu2024kivi}. This suggests that reorganizing floating-point data in a manner that exploits channel-wise similarity could unlock better compressibility. 

Although these analyses typically target model interpretability or activation sparsity, they illuminate the possibility of \emph{reformatting} the data layout to expose additional redundancy to compressors like LZ4/ZSTD. For example, exponent values along the same channel across multiple tokens may follow narrow numeric distributions or repeat patterns, offering a fertile path for lossless compression methods. In the remainder of this work, we investigate how such insights can be used to systematically reorganize floating-point weights and KV caches, enabling conventional compressors to achieve more significant savings without sacrificing any inference accuracy.

\subsection{Dynamic Quantization}
Given the limitations of fixed-precision memory access, dynamic quantization has emerged as a flexible strategy to manage precision based on contextual importance, enhancing memory latency and throughput. Dynamic quantization in KV cache applies adaptive precision based on the relevance of each token’s embedding, allowing high-importance tokens to retain higher precision (e.g., BF16, FP16), while less critical tokens are quantized to lower precisions (e.g., FP8 or FP4). As shown in Table~\ref{tab: dynamic quantization perplexity}, dynamic quantization on the LLaMA 3.1 8B model with the BookSum dataset maintains perplexity close to that of a full KV cache setup. For instance, quantizing the top 5 pages to BF16 (a page contains 16 tokens), the next 5 to FP8, and the next 3 to FP4 yields a perplexity of 11.60, compared to 10.49 for the full KV cache configuration and 12.49 in the Quest setup~\cite{tang2024quest}.

\begin{table}[]
\centering
\caption{Perplexity for various quantization methods on LLaMA 3.1 8B model and BookSum dataset}
\label{tab: dynamic quantization perplexity}
\resizebox{\columnwidth}{!}{%
\begin{tabular}{@{}ll@{}}
\toprule
\textbf{Method}                                                          & \textbf{Perplexity} \\ \midrule
Full KV Cache                                                            & 10.49               \\
Sliding Window (64 tokens)                                               & 14.33               \\
Quest (Top 5 pages in BF16)                                              & 12.49               \\
Dynamic Quant. (Top 5 pages in BF16, Next 3 in FP8, Next 2 in FP4) & 11.87               \\
Dynamic Quant. (Top 5 pages in BF16, Next 5 in FP8) & 11.60               \\ \bottomrule
\end{tabular}%
}
\end{table}

\begin{figure}[htbp]
	\centering
    \includegraphics[width=\linewidth]{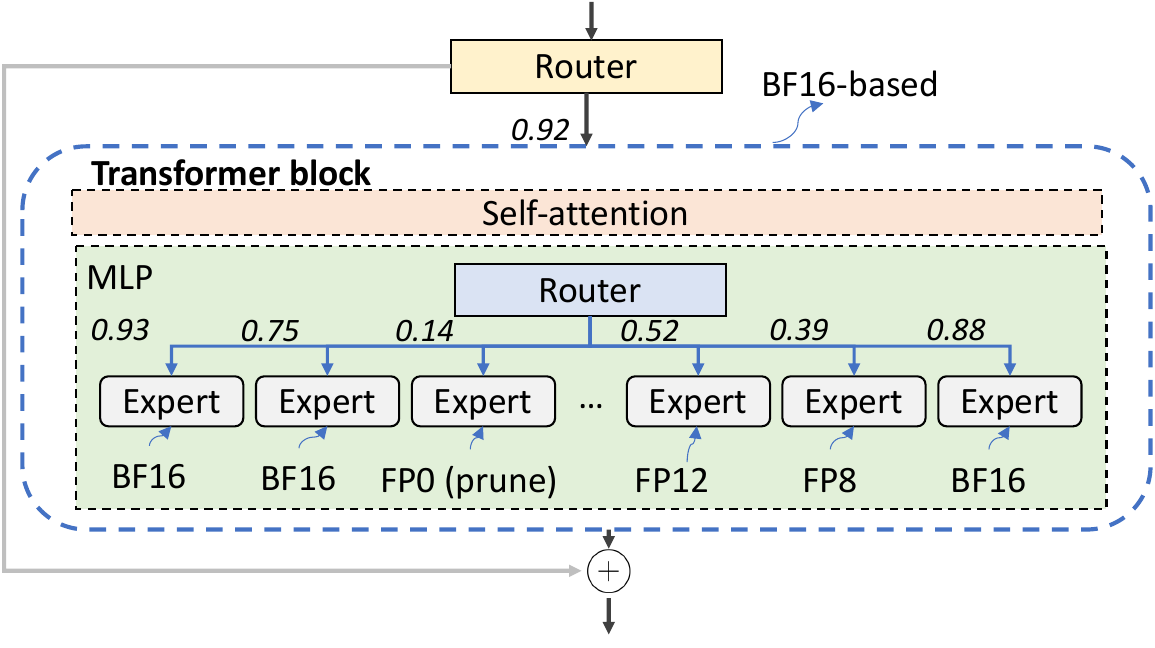}
	\caption{Illustration of dynamic weight quantization in a transformer block based on Mixture-of-Depth-Expert (MoDE)~\cite{raposo2024mixture}.}
\label{fig.illustration MoDE dynamic quantization}
\end{figure}

\begin{figure}[htbp]
	\centering
    \includegraphics[width=\linewidth]{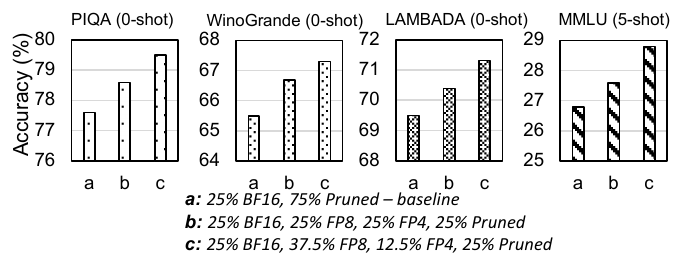}
	\caption{Accuracy comparison among quantization configurations on prune-only (a) and dynamic quantization (b), (c) based on LLaMA-MoE-3.5B~\cite{zhu2024llama} on PIQA~\cite{bisk2020piqa}, WinoGrande~\cite{sakaguchi2021winogrande}, LAMBADA~\cite{paperno2016lambada} and MMLU~\cite{hendrycks2020measuring} datasets.}
\label{fig.moe dynamic quantization}
\end{figure}

Dynamic quantization also extends to model weights, where precision adjustments based on contextual relevance allow finer granularity than dynamic pruning. Unlike traditional pruning methods, which only activate or deactivate model weights, dynamic quantization allows real-time precision adjustments (e.g., FP16, FP8, FP4), refining memory and energy efficiency by tuning precision to match computational needs. Within the Mixture of Depths and Experts (MoDE) framework~\cite{raposo2024mixture}, we implemented a context-dependent dynamic weight quantization, where routers control the precision level for each component of the model block, as illustrated in Fig.~\ref{fig.illustration MoDE dynamic quantization}. A performance comparison on the LLaMA-MoE-3.5B model (Fig.~\ref{fig.moe dynamic quantization}) demonstrates the effectiveness of dynamic quantization in balancing memory efficiency and task accuracy across several configurations. For instance, a configuration quantizing more experts to lower precisions rather than simply skipping those experts achieved an improvement in zero-shot accuracy on the PIQA task, with a 1.9 percentage points increase over the baseline configuration. These results validate dynamic quantization as a robust method to enhance memory efficiency while sustaining performance across varied precision settings.

Modern AI accelerators are equipped with hardware support for variable-precision arithmetic, enabling them to effectively utilize dynamic quantization to improve computational efficiency. However, because dynamic quantization is implemented within the accelerators themselves, its ability to proportionally reduce memory bandwidth and energy consumption remains unclear.

\section{Proposed Design}\label{sec:design}
We developed a design solution that enables AI accelerators to better address memory bottlenecks by enhancing their on-chip memory controllers. This solution significantly improves the compressibility of in-memory data and ensures that memory bandwidth/energy consumption scale gracefully with dynamic quantization. It contains two key design techniques:
\begin{enumerate}
\item \textit{Bit-plane disaggregation}: 
This method decomposes model weights and KV cache data into individual \emph{bit-planes}, yielding notably higher lossless compressibility and reducing memory bandwidth consumption when dynamic quantization is applied. By storing exponent and mantissa bits separately in contiguous planes, classical block compressors (e.g., LZ4, ZSTD) can more easily detect repeating patterns, and partial-plane fetching becomes possible.

\item \textit{Cross-token KV cache clustering and de-correlation}: This technique aims at further improving KV cache compressibility. Motivated by the observation that data points in the same position across tokens often exhibit high similarity~\cite{liu2024kivi}, we organize the KV cache by grouping numerical data points at the same positions across tokens (i.e., channel-wise grouping by specific heads and embedding dimensions across tokens). In addition, we apply content de-correlation (e.g., subtraction or bit-wise XOR) to further improve the lossless compressibility.
\end{enumerate}

\begin{figure}[htbp]
	\centering
    \includegraphics[width=\linewidth]{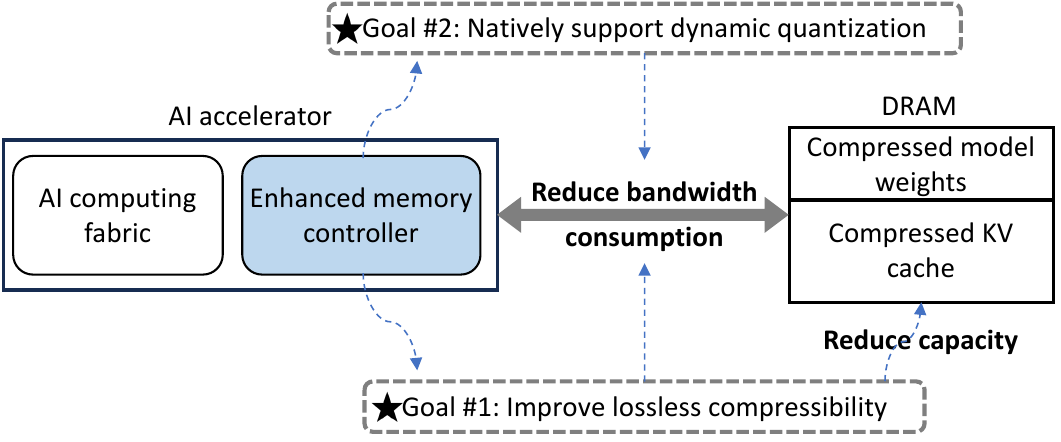}
	\caption{Mitigating memory bottlenecks by enhancing on-chip memory controller within AI accelerators.}
\label{fig.architecture overview}
\end{figure}

To ensure practical feasibility, we integrate these techniques—along with a hardware-based (de)compression engine—within the on-chip memory controller, as illustrated in Fig.~\ref{fig.architecture overview}. This approach requires minimal changes to both the computational framework and software stack, as the memory controller merely needs to recognize whether data are weights or KV caches. All data rearrangements and compression steps occur entirely in hardware at the memory controller, remaining transparent to the accelerator’s computing fabric and the broader software stack. The next subsections further detail how bit-plane disaggregation and cross-token KV cache clustering operate in practice.

\subsection{Bit-plane Disaggregation}

Floating-point data (e.g., FP16) exhibits low compressibility when stored in the conventional per-number layout. Each 16-bit value commingles exponent and fraction bits in a way that obscures repeated patterns. Hence, the concept of \emph{bit-plane disaggregation} is introduced, which reorganizes these floating-point numbers according to bit position. Critically, exponents from different values become co-located in one plane, which can enable classical lossless compressors (e.g., LZ4, ZSTD) to exploit their typically low entropy. 

Consider an $n$-bit floating-point format, partitioned as $1$ sign bit $s_j$, $E$ exponent bits $e_{j,1\ldots E}$, and $F$ fraction (mantissa) bits $f_{j,1\ldots F}$, with $n = 1 + E + F$. Thus each floating-point number $x_j$ can be written in IEEE-like notation as 
\begin{equation}
   x_j \;=\; (-1)^{s_j} \,\times\, 2^{\,(\mathrm{Exp}(e_j)-\mathrm{Bias})} \,\times\, \left(1 + \mathrm{Frac}(f_j)\right),
\end{equation}
where $\mathrm{Exp}(e_j)$ and $\mathrm{Frac}(f_j)$ decode the exponent and fraction fields, and $\mathrm{Bias}$ is the usual offset for the exponent. Although the internal decoding involves exponent/fraction arithmetic, at the bit level we can treat $x_j$ as a simple array of $n$ bits:
$x_j \;\longleftrightarrow\; \bigl[s_{j},\, e_{j,1},\ldots,e_{j,E},\, f_{j,1},\ldots,f_{j,F}\bigr]$.
Let us denote these bits by $b_{j,n-1}$ down to $b_{j,0}$, from the most significant to the least significant.

\begin{figure}[htbp]
	\centering
    \includegraphics[width=\linewidth]{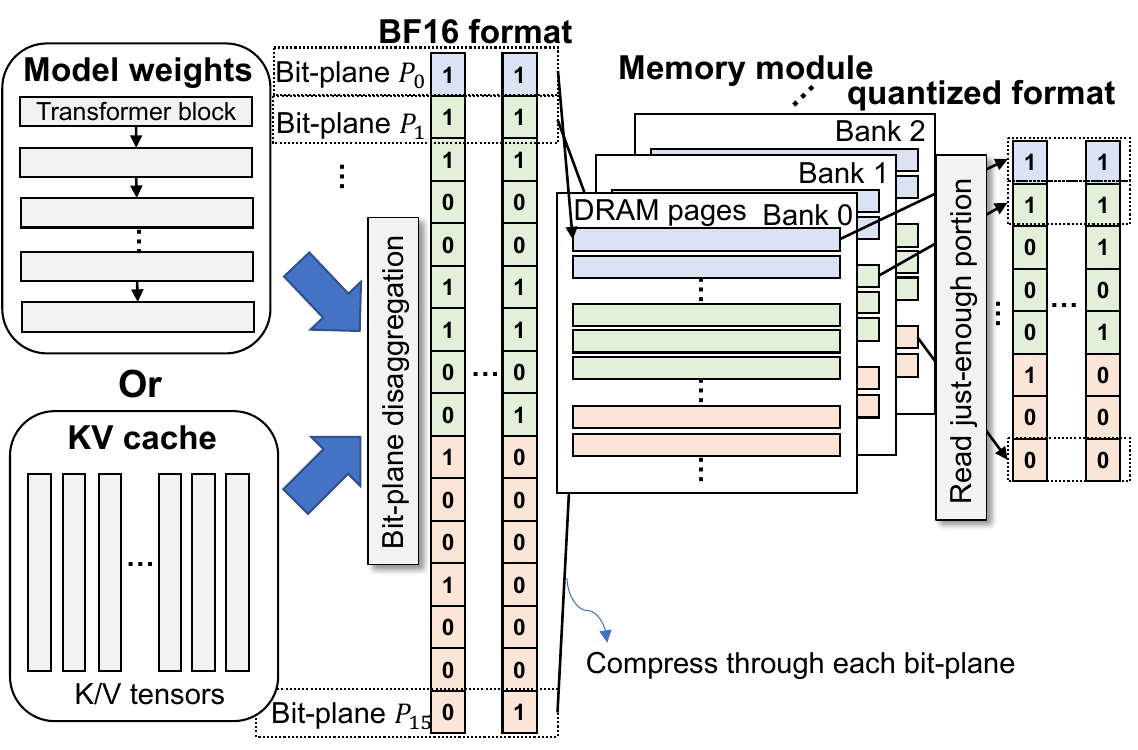}
	\caption{Illustration of \textit{bit-plane disaggregation} in-memory placement.}
\label{fig.bitplane placement}
\end{figure}

Now, suppose we have a block of $m$ such $n$-bit floating-point values $\{\, x_1,\,x_2,\,\ldots,x_m \}$.
As shown in Fig.~\ref{fig.bitplane placement}, we define the \( i \)-th bit-plane \( P_i \) as the set of all \( i \)-th bits across the entire collection of data points $P_i = \{ b_{1,i}, b_{2,i}, \dots, b_{m,i} \}$, 
where \( i \) ranges from \( 0 \) to \( n-1 \). Thus, each bit-plane \( P_i \) contains the bits from all data points at position \( i \), effectively creating a "slice" through the data based on bit significance.
The complete set of bit-planes for the data collection can be represented as a matrix \( \mathbf{P} \), where each row corresponds to a bit-plane:
\begin{equation}
\mathbf{P} = 
\begin{bmatrix}
P_0 \\
P_1 \\
\vdots \\
P_{n-1}
\end{bmatrix} =
\begin{bmatrix}
b_{1,0} & b_{2,0} & \dots & b_{m,0} \\
b_{1,1} & b_{2,1} & \dots & b_{m,1} \\
\vdots & \vdots & \ddots & \vdots \\
b_{1,n-1} & b_{2,n-1} & \dots & b_{m,n-1}
\end{bmatrix}
\label{equ.bitplane represent}
\end{equation}
In this matrix, each row \( P_i \) represents a bit-plane, grouping bits by their position across all data points. This disaggregation into bit-planes enables tailored compression for each plane, as each bit-plane has different characteristics. For example, higher-order bit-planes (representing the most significant bits) typically exhibit lower entropy and therefore greater redundancy, making them more suitable for aggressive compression.

We note that this reorganized layout also enables \emph{selective retrieval} of high-order bit-planes for dynamic quantization as shown in Fig.~\ref{fig.bitplane placement}. For example, if we decide to reduce FP16 data to just the top 8 bits in certain layers, we can read only bit-planes $8\ldots 15$ from memory and skip the remainder, reducing bandwidth demand without altering the hardware’s arithmetic capabilities.


From a hardware standpoint, the memory controller maintains a small \emph{bit-plane aggregator} module to perform this disaggregation. Internally, a crossbar or shuffle network routes input bits to plane-specific buffers (usually sized at 1--4\,KB). Once a full block is assembled, an on-chip compression engine (LZ4, ZSTD, or custom IP) compresses each bit-plane, storing optional metadata (e.g., partial-plane indices) in a compact header. During retrieval, the controller fetches and decompresses only the necessary bit-planes (e.g., top 8 planes for FP8) and reconstitutes the data into standard floating-point layout before sending it to the compute fabric. This organization \emph{naturally} supports dynamic quantization by allowing selective omission of lower-order planes, thereby reducing DRAM traffic in direct proportion to the chosen precision. 

\subsection{Cross-Token KV Cache Clustering and De-correlation}

Cross-token KV cache clustering and de-correlation enhance compressibility by grouping KV cache tensors across multiple tokens and organizing them based on positional alignment. For simplicity, we represent each token’s KV tensor as a vector, denoted by \( \mathbf{k}_t \) for token \( t \), where each position in \( \mathbf{k}_t \) corresponds to an embedding dimension. By aligning these vectors within a group of \( n \) tokens, we create a matrix structure \( G_{j} \) that captures redundancy more effectively. Grouping KV entries in this way allows for more efficient memory usage by handling data at a group level, improving organization and compressibility.

\subsubsection{Channel-Wise Grouping Across Tokens}

\begin{figure}[htbp]
	\centering
    \includegraphics[width=\linewidth]{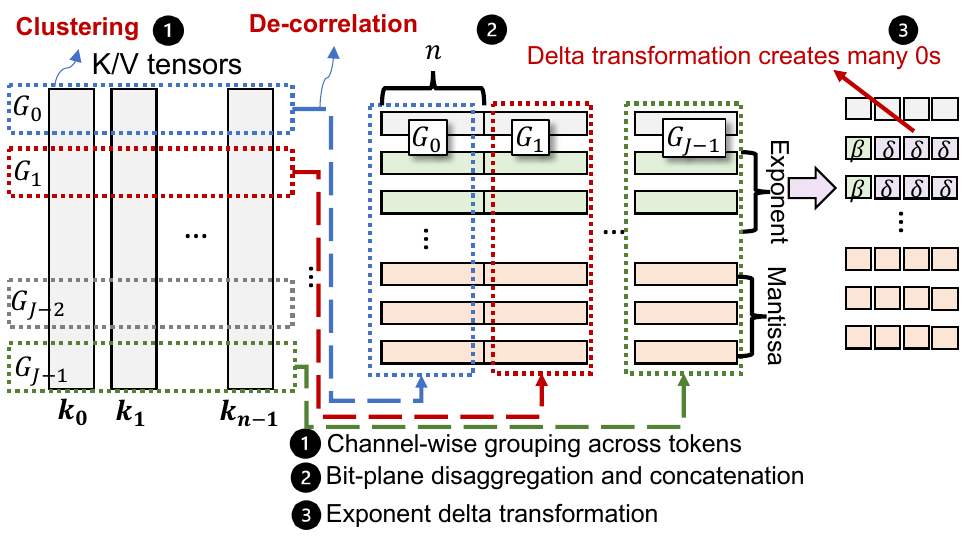}
    \caption{Illustration of cross-token KV cache clustering and de-correlation for a group of \( n \) tokens, showing channel-wise grouping across tokens, bit-plane disaggregation and concatenation and exponent delta transformation.}
\label{fig.cross-token kv}
\end{figure}

As shown in Fig.~\ref{fig.cross-token kv} \blackcircle{1}, in each token group \( \mathbf{G} \), we organize KV vectors \( \mathbf{k}_t \) by aligning entries at the same position across all tokens in the group, thereby forming a matrix structure for each channel. For each channel \( j \), representing a specific entry within each token’s KV vector, we collect the entries at channel \( j \) across all tokens in the group:
\begin{equation}
G_{j} = \{ k_{t, j} \mid t = 0, \dots, n-1 \}.
\end{equation}
Here, each \( k_{t, j} \) represents the entry at channel \( j \) in token \( t \)’s KV vector \( \mathbf{k}_t \), and \( G_{j} \) becomes a row of entries aligned by channel across tokens within the group \( G \). This structure enhances compressibility by aligning similar data elements.

\subsubsection{Bit-Plane Disaggregation and Concatenation on KV Cache}

To enhance compressibility, we begin by organizing each entry in \( G_{j} \) into bit-planes, isolating each bit position across all tokens in a group, as shown in Fig.~\ref{fig.cross-token kv} \blackcircle{2}. Each KV entry \( k_{t, j} \) is represented as a binary sequence, and the \( i \)-th bit-plane \( P_i(G_{j}) \) for \( G_{j} \) (structured as a matrix) is:
\begin{equation}
P_i(G_{j}) = \{ \text{Bit}(k_{t, j}, i) \mid t = 0, \dots, n-1 \}.
\end{equation}
Once all bit-planes are extracted, we concatenate these bit-planes across all positions \( j \) within each group to form a single bit-plane sequence, as illustrated in Fig.~\ref{fig.cross-token kv} \blackcircle{2}:
\begin{equation}
\text{Concatenated\_Bitplane}(G_{i}) = \bigcup_{j=0}^{J-1} P_i(G_{j}),
\end{equation}
where \( J \) is the number of positions within the KV vectors. 

\subsubsection{Exponent Delta Transformation}

Following bit-plane disaggregation, we apply an exponent delta transformation to reduce the range of exponent values. For each token group, we identify a base exponent \( \beta_{j} \) — the minimum or most common exponent across all tokens for position \( j \). Each exponent in \( G_{j} \) is then transformed relative to \( \beta_{j} \), as shown in Fig.~\ref{fig.cross-token kv} \blackcircle{3}. The transformed exponent for each KV entry \( k_{t, j} \) is then:
\begin{equation}
\delta_{t, j} = \text{Exponent}(k_{t, j}) - \beta_{j}.
\end{equation}
The delta-transformed form of \( G_{j} \) then consists of the base exponent \( \beta_{j} \) and the transformed delta values:
\begin{equation}
G_{j} = \{\beta_{j}, \delta_{t, j} \mid t = 0, \dots, n-1 \}.
\end{equation}
Finally, the delta-transformed bit-planes are packed into bytes and compressed, optimizing memory usage and enabling efficient data retrieval during inference.

From a hardware perspective, the memory controller includes a dedicated \emph{channel-wise KV aggregator} module that buffers token embeddings (e.g., a batch of \( n \) tokens) and rearranges them so that channel \( j \) values appear contiguously. A small integer subtractor computes the exponent delta \(\delta_{t,j}\) relative to \(\beta_{j}\), which is stored in a per-channel metadata buffer. The bit-plane shuffle network then disaggregates \(\delta_{t,j}\) and mantissa bits into planes, similar to the model-weight path. Finally, an on-chip block compression engine (LZ4/ZSTD) encodes the rearranged KV data. During reads, the controller reverses these steps by decompressing the bit-planes, restoring exponents via \(\beta_{j}\)+\(\delta_{t,j}\), and outputting the original per-token KV layout. In practice, additional small header fields (one base exponent per channel) are stored with each block. 

\section{Evaluation}\label{sec:experiment}

\subsection{Compression Efficiency}

\subsubsection{KV Cache Compressibility}
\begin{figure*}[htbp]
	\centering
    \includegraphics[width=\linewidth]{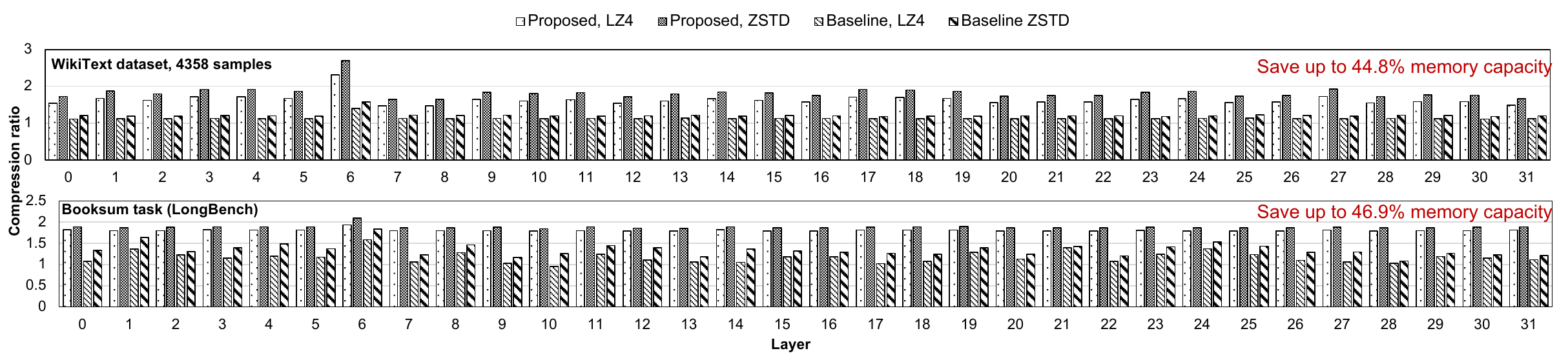}
	\caption{Comparison of KV cache bit-plane 4KB compression ratios in the LLaMA 3.1 8B model (32 layers) using WikiText (test) datasets and BookSum task (LongBench). We evaluate \textit{cross-token KV cache clustering and de-correlation} versus \textit{baseline approach} with LZ4 and ZSTD compression with 4KB compression block.}
\label{fig.kv cache by layer compression ratio}
\end{figure*}

We evaluated the compressibility of the KV cache across 32 layers of the LLaMA 3.1 8B model on the WikiText dataset~\cite{wikitext-website} and BookSum (LongBench) datasets~\cite{kryscinski2021booksum}. In this work, we define {\it compression ratio} as ${S_{orig}}/{S_{comp}}\ge 1$, where $S_{orig}$ and $S_{comp}$ denote the size of the original and compressed data blocks. Compression ratios were measured for both LZ4 and ZSTD algorithms with a 4KB compression block size, as shown in Fig.\ref{fig.kv cache by layer compression ratio}. Our data placement strategy on the KV cache achieved 44.8\% and 46.9\%  overall footprint reduction on WikiText and Booksum task. On WikiText, the highest compression ratios on a single layer reached 2.69 (ZSTD) and 2.31 (LZ4), while on BookSum, they peaked at 2.10 (ZSTD) and 1.93 (LZ4). Compared to the baseline (with overall ZSTD compression ratios of 1.21 on WikiText and 1.33 on BookSum), which does not apply \textit{cross-token KV cache clustering and de-correlation}, our approach (with overall ratios of 1.81 on WikiText and 1.88 on BookSum) improves the overall KV cache lossless compression ratio by 50.3\% on WikiText and 41.7\% on BookSum using ZSTD.

\subsubsection{Model Weights Compressibility}

Table~\ref{tab:model weight compression ratio} presents compression ratios and corresponding memory savings for various LLM configurations across different precision levels on model weights. BF16-based models achieve the highest lossless compression gains; for instance, LLaMA 3.1 8B in BF16 precision achieves a ZSTD lossless compression ratio of 1.34, leading to a 25.2\% reduction. Moreover, since the proposed method is orthogonal to recent lossy compression techniques, it can also cooperate effectively with quantization approaches (e.g., GPTQ~\cite{frantar2022gptq}) to amplify total memory savings. For example, for FP8 and INT4 precision models, combining our method with AutoFP8 and GPTQ lossy compression—starting from BF16—results in substantial overall savings. For instance, quantizing LLaMA 3.1 8B to FP8 achieves a 54.1\% total reduction, merging a 50\% lossy saving with an additional 8.3\% from our lossless compression approach.

\begin{table}[htbp]
\centering
\caption{Lossless compression ratios and total memory savings with lossy compression on model weights}
\label{tab:model weight compression ratio}
\resizebox{\linewidth}{!}{%
\begin{tabular}{@{}lllll@{}}
\toprule
Model          & Precision & Comp. Ratio & Lossless Savings & Total Savings \\ \midrule
LLaMA 3.1 8B   & BF16      & 1.34              & 25.2\%          & 25.2\%      \\
               & FP8       & 1.09              & 8.3\%           & 54.1\%      \\
               & INT4      & 1.01              & 0.9\%           & 75.2\%      \\
LLaMA 3.1 70B  & BF16      & 1.34              & 25.6\%          & 25.6\%      \\
               & FP8       & 1.10              & 9.3\%           & 54.6\%      \\
               & INT4      & 1.02              & 2.1\%           & 75.5\%      \\
Mixtral 8×7B   & BF16      & 1.32              & 24.4\%          & 24.4\%      \\
               & FP8       & 1.09              & 8.0\%           & 54.1\%      \\
               & INT4      & 1.01              & 1.2\%           & 75.3\%      \\
LLaMA MoE 3.5B & BF16      & 1.33              & 24.9\%          & 24.9\%      \\
               & FP8       & 1.11              & 9.9\%           & 54.9\%      \\
               & INT4      & 1.02              & 1.6\%           & 75.4\%      \\ \bottomrule
\end{tabular}%
}
\end{table}

\begin{figure}[htbp]
	\centering
    \includegraphics[width=\linewidth]{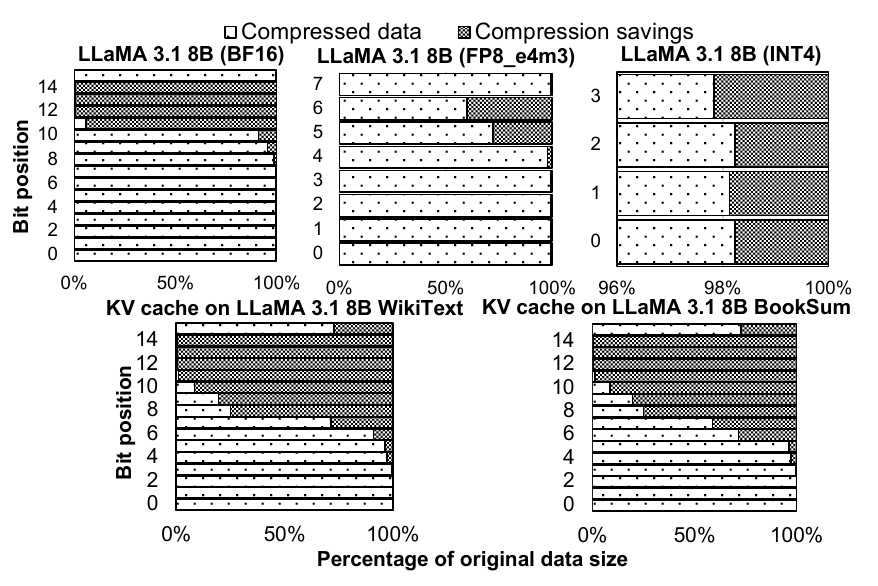}
	\caption{Compressibility of model weights and KV cache bit-planes in the LLaMA 3.1 8B model, evaluated for BF16, FP8, and INT4 weight formats and BF16 KV cache on WikiText and BookSum datasets, utilizing ZSTD compression with 4KB blocks.}
\label{fig.model parameter compressibility}
\end{figure}

Fig.~\ref{fig.model parameter compressibility} illustrates the compressibility of model weights across bit-planes for BF16, FP8, and INT4-based LLMs using 4KB ZSTD lossless compression, along with KV cache compressibility in the LLaMA 3.1 8B model on WikiText and BookSum datasets. For BF16 model weights, the top four exponent bit-planes contribute the most to overall compressibility, and together with other bit-planes it achieves an overall compression ratio of 1.34. This is because exponents, especially in high-precision formats like BF16, often contain more redundancy and fewer unique values, allowing for effective compression. In contrast, FP8 and INT4 models, already subjected to lossy quantization, show limited compressibility as the reduced bit precision minimizes representational redundancy, particularly in the exponent bits.

For the KV cache in BF16 format, shown in the lower two subfigures of Fig.~\ref{fig.model parameter compressibility}, the exponent bit-planes again demonstrate significantly higher compressibility. This is attributed to the relatively narrow range of data stored along the channel in KV cache, where exponents frequently exhibit low variability across tokens. These properties lead to substantial memory savings of 44.8\% on the WikiText and 46.9\% on BookSum (as shown in Fig.~\ref{fig.kv cache by layer compression ratio}).

\begin{figure*}[htbp]
	\centering
    \includegraphics[width=\linewidth]{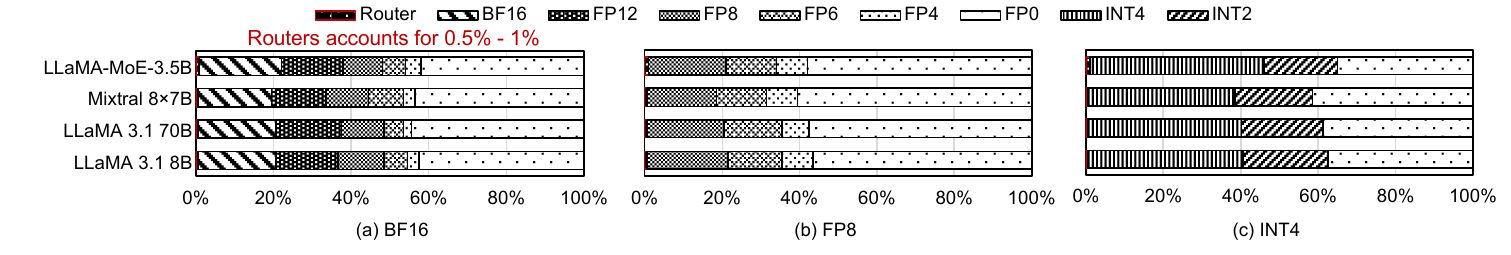}
	\caption{Precision distribution for model weights of LLaMA 3.1 8B, LLaMA 3.1 70B, Mixtral 8×7B, and LLaMA-MoE-3.5B in BF16, quantized FP8, and INT4 when conducting inference on WikiText-2.}
\label{fig.precision_distribution}
\end{figure*}

\subsection{DRAM Access Efficiency with Dynamic Quantization}
Using the open-source DRAM simulator DRAMSim3~\cite{li2020dramsim3}, we further evaluated the DRAM access efficiency. We set that each memory module contains 4 DRAM channels, each channel hosting 10 ×4 DDR5-4800 devices. 

To facilitate dynamic quantization, we adapted the LLaMA 3.1 models—specifically the 8B and 70B variants~\cite{dubey2024llama}, as well as the Mixtral 8×7B model~\cite{jiang2024mixtral}, into MoDE architecture~\cite{raposo2024mixture} as depicted in Fig.~\ref{fig.moe dynamic quantization}. Additionally, we transformed the dense MLP layers in LLaMA into a MoE structure to enhance flexibility for dynamic quantization. To mitigate fine-tuning costs, we applied Low-Rank Adaptation (LoRA)~\cite{hu2021lora}, utilizing the C4 dataset to calibrate the router parameters. Subsequently, we employed AutoFP8~\cite{autofp8} and GPTQ~\cite{frantar2022gptq} techniques to obtain models quantized to FP8 and INT4 precisions by using Ultrachat~\cite{ding2023enhancing} as calibration dataset. 

Fig.~\ref{fig.precision_distribution} shows 12 LLMs in aforementioned configuration and their average model weights precision distribution when conducting inference on WikiText-2~\cite{wikitext-website}. The inference were configured with various dynamic quantization schemes: BF16, FP12/8/6/4 applied to BF16-based models; FP8/6/4 applied to FP8-based models; and INT4/2 applied to INT4-based models. All router layers are using BF16 precision for accuracy. 

\begin{figure}[htbp]
	\centering
    \includegraphics[width=\linewidth]{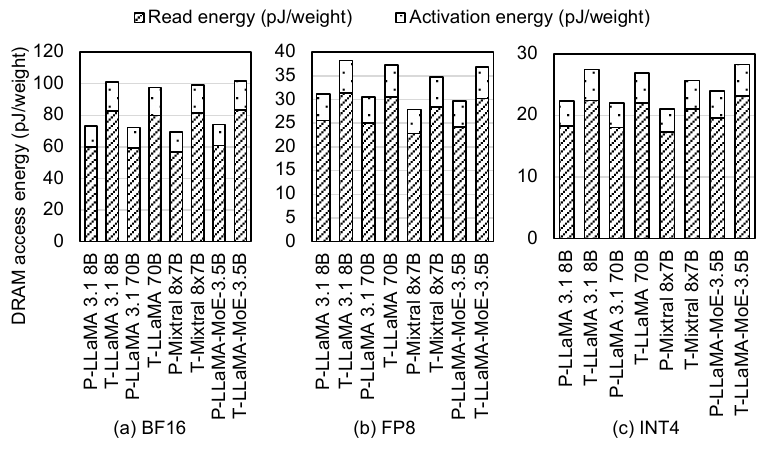}
	\caption{DRAM access energy per weight for models LLaMA 3.1 8B, LLaMA 3.1 70B, Mixtral 8×7B, and LLaMA-MoE-3.5B in BF16, quantized FP8, and INT4 when conducting inference on WikiText-2. We compared read and activation energy for Proposed bit-plane (P) and Traditional byte-level (T) approach.}
\label{fig.access_energy}
\end{figure}

Fig.~\ref{fig.access_energy} compares the DRAM access energy with dynamic quantization, and shows read and activation energy between the proposed bit-plane method (P) and a traditional straightforward byte-level approach (T). Implementing the proposed method results in a DRAM access energy reduction of up to 29.9\%. Specifically, for BF16-based models, the proposed method achieves energy reductions of 27.8\% for LLaMA 3.1 8B, 25.9\% for LLaMA 3.1 70B, 29.9\% for Mixtral 8×7B, and 27.2\% for LLaMA MoE 3.5B. In the case of the Mixtral 8×7B model, the proposed method reduces DRAM access energy by 19.6\% for the FP8-based model and 17.9\% for the INT4-based model compared to the traditional approach, which shows a decreasing trend in energy savings as the quantization precision of the model decreases.


\begin{figure}[htbp]
	\centering
    \includegraphics[width=\linewidth]{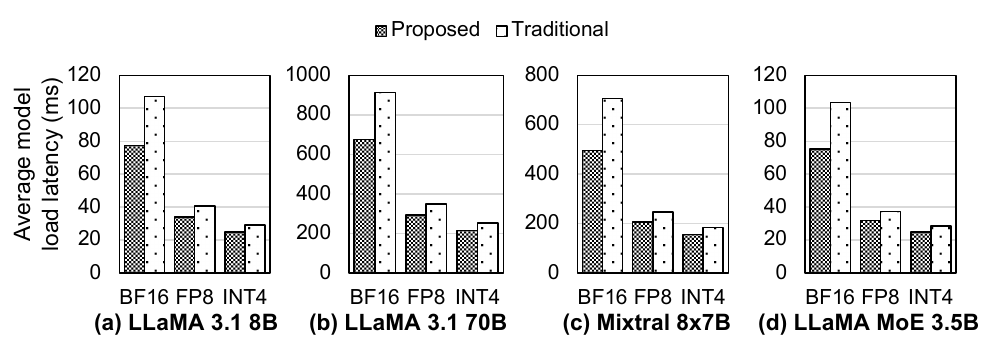}
	\caption{Average model load latency for models LLaMA 3.1 8B, LLaMA 3.1 70B, Mixtral 8×7B, and LLaMA-MoE-3.5B in BF16, quantized FP8, and INT4 when conducting inference on WikiText-2.}
\label{fig.model load latency}
\end{figure}

Fig.~\ref{fig.model load latency} presents the average model load latency on the WikiText-2 dataset for aforementioned 12 LLMs when applying same dynamic quantization.
The proposed method achieves latency reductions of up to 30.0\%. In Mixtral 8×7B, the BF16 configuration shows a decrease from 705.90 to 495.06 ms (a 30.0\% reduction), while the FP8 and INT4 formats achieve respective reductions of approximately 17.1\% and 14.5\%. For the larger LLaMA 3.1 70B model, BF16 latency is cut from 910.58 to 674.73 ms (25.9\% reduction), with the FP8 and INT4 versions also exhibiting notable improvements (from 348.65 to 293.27 ms and from 251.03 to 214.11 ms, respectively). These results demonstrate that the proposed design not only minimizes DRAM traffic but also significantly reduces load latency across various architectures and precision levels, thereby boosting overall inference efficiency.

\begin{table*}[htbp]
    \centering
    \caption{Silicon cost at 2\,GHz with 32 lanes for LZ4 and ZSTD lossless compression. (``SL'' denotes single-lane.)}
    \label{tab:simulation}
    \begin{tabular}{lrrrrrr}
    \toprule
    \textbf{Engine} & \textbf{BlockSize} & \textbf{SL Area} & \textbf{SL Power} & \textbf{LaneTotArea} & \textbf{LaneTotPower} & \textbf{SL Thpt} \\
     & (bits) & (mm\(^2\)) & (mW) & (mm\(^2\)) & (mW) & (Gbps) \\
    \midrule
    LZ4  & 16384 & 0.05669 & 696.515  & 1.81413 & 2228.846 & 512 \\
    LZ4  & 32768 & 0.07557 & 885.258  & 2.41811 & 2832.826 & 512 \\
    LZ4  & 65536 & 0.15106 & 1640.233 & 4.83403 & 5248.745 & 512 \\
    \midrule
    ZSTD & 16384 & 0.08357 & 1363.715 & 2.67429 & 4363.886 & 512 \\
    ZSTD & 32768 & 0.10245 & 1552.458 & 3.27827 & 4967.866 & 512 \\
    ZSTD & 65536 & 0.17794 & 2307.433 & 5.69419 & 7384.785 & 512 \\
    \bottomrule
    \end{tabular}
\end{table*}

\subsection{Hardware Implementation and Resource Evaluation}
We implemented a parameterizable RTL design of our bit-plane–aware compression subsystem in SystemVerilog, including several modules: (1) Bit-plane aggregator: responsible for shuffling data into bit-planes (and de-shuffling upon decompression); (2) Compression engine supporting both encoding and decoding; (3) Control logic and buffers: handling block-based input/output (e.g., 2KB or 4KB), scheduling read/write requests, and interfacing with the on-chip memory controller.

Table~\ref{tab:simulation} summarizes the hardware resource usage at 2\,GHz with 32 parallel lanes for both LZ4 and ZSTD compression engines (synthesized with 7\,nm PDKs~\cite{clark2016asap7}), evaluated over three block sizes (16384, 32768, and 65536 bits). For LZ4, the single lane area increases from 0.057\,mm\(^2\) at 16384 bits to 0.151\,mm\(^2\) at 65536 bits, while the Single-Lane Power increases from 696.515\,mW to 1640.233\,mW. For ZSTD, the Single-Lane Area increases from 0.084\,mm\(^2\) to 0.178\,mm\(^2\), and the Single-Lane Power rises from 1363.715\,mW to 2307.433\,mW. Aggregating across 32 lanes, LZ4 achieves total areas of 1.814\,mm\(^2\) (16384 bits), 2.418\,mm\(^2\) (32768 bits), and 4.834\,mm\(^2\) (65536 bits), with corresponding total powers of 2228.846\,mW, 2832.826\,mW, and 5248.745\,mW. Similarly, ZSTD achieves total areas of 2.674\,mm\(^2\), 3.278\,mm\(^2\), and 5.694\,mm\(^2\) with total powers of 4363.886\,mW, 4967.866\,mW, and 7384.785\,mW, respectively. Each lane delivers a throughput of 512\,Gbps, resulting in an aggregate throughput of 16384\,Gbps (2 TB/s) for 32 lanes.

\section{Related Works}

\noindent\textbf{Bit-Plane Disaggregation}: The use of bit-plane disaggregation for improving data compressibility and hardware efficiency has been explored extensively in both classical and contemporary literature. Early works such as BPC \cite{kim2016bit} by NVIDIA highlight that regrouping bits in a bit-plane manner can notably enhance the compressibility of uniformly typed data blocks, which offers a more hardware-friendly implementation compared to traditional byte- or word-oriented layouts. EBPC \cite{cavigelli2019ebpcextendedbitplanecompression} applied an extended bit-plane compression scheme to deep neural network accelerators, which demonstrates higher compression ratios by taking advantage of bit-level redundancy within activations. More recent approaches extend these concepts specifically to LLMs and other deep networks with mixed-precision arithmetic. For example, \cite{park2024any} organizes quantized parameters into distinct bit-planes to facilitate “any-precision” execution — that is, it allows loading only the necessary precision based on dynamic accuracy-speed trade-offs. In a similar vein, SmartQuant \cite{xie2024smartquant} uses bit-plane techniques to store LLM weights in a partially quantized format, dynamically retrieving different subsets of bit-planes according to the context’s numerical requirements. 

\noindent\textbf{Lossy Compression in LLMs}: Post-training quantization approaches, such as GPTQ~\cite{frantar2022gptq}, AWQ~\cite{lin2024awq}, RPTQ~\cite{yuan2023rptq} and SmoothQuant~\cite{xiao2023smoothquant}, have been developed to convert model weights and activations to lower bit-width representation, effectively reducing model size and inference latency. SparseGPT~\cite{frantar2023sparsegpt} applies structured pruning to LLMs, effectively reducing model size and computational requirements. Spqr~\cite{dettmers2023spqr} explores sparse quantization, combining pruning and quantization to enhance efficiency. 

\noindent\textbf{Contextual Importance in LLMs}: Deja Vu~\cite{liu2023deja} is a framework that predicts contextual sparsity on-the-fly for each input. PowerInfer~\cite{song2023powerinfer}, LLM in a flash~\cite{alizadeh2023llm} extend this work to hybrid CPU/GPU and flash platforms. MoE and MoD~\cite{raposo2024mixture} dynamically adjust the depth and experts across different tokens or layer within a model based on the input tokens.

\noindent\textbf{Lossless Memory Compression}: Multiple approches of implementation of hardware-based main memory compression has been proposed~\cite{choukse2018compresso,ekman2005robust,pekhimenko2013linearly,zhao2015buri}. Key-value store system like ZipCache~\cite{xie2024zipcache}, ZipKV~\cite{ma2023zipkv} apply memory block compression to the DRAM tier to reduce the memory footprint. Contemporary data analytics systems like SAP HANA~\cite{farber2012sap}, Oracle~\cite{lahiri2015oracle}, and Snowflake~\cite{dageville2016snowflake} apply block compression to reduce their memory consumption. 

\section{Conclusion}

This paper has presented a memory controller–centric solution for mitigating the bandwidth and capacity bottlenecks that commonly arise in LLM inference. By judiciously reorganizing floating-point data at the bit level and exploiting cross-token correlation in the KV cache, our design enables efficient application of standard lossless compressors (LZ4, ZSTD). Experimental results on a range of publicly available LLMs demonstrate that this approach reduces model weights by up to 25.2\% and KV cache by as much as 46.9\%—all without degrading model accuracy. Moreover, the proposed layout allows dynamic quantization to scale memory bandwidth and energy consumption proportionally, achieving up to 30.0\% faster model load times and 29.9\% lower DRAM access energy in DRAMSim3 simulations. Our hardware evaluation further reveals that even when instantiated at 2\,GHz across 32 parallel lanes, the total area overhead remains modest (e.g., 4.83\,mm\(^2\) for an LZ4-based design and 5.69\,mm\(^2\) for ZSTD), while delivering up to 2\,TB/s of effective throughput. These findings underscore the pivotal role that an LLM-aware memory controller can play in meeting the growing performance and efficiency demands of modern AI inference workloads.


\bibliographystyle{IEEEtran}
\bibliography{references}


\end{document}